\documentclass[preprint]{raa}           
\usepackage{graphicx,times}
\usepackage{subfigure}
\usepackage{natbib}
\usepackage{amssymb,amsmath}
\bibpunct{(}{)}{;}{a}{}{,}

\newcommand{\HI}{$\mathrm{HI}$\,} 

\begin{document}

\title{Sky reconstruction for the Tianlai cylinder array}

\author{Jiao Zhang\inst{1,2,3}, Shifan Zuo\inst{1,3},  Reza Ansari\inst{2}, Xuelei Chen\inst{1,3,4} ,Yichao Li\inst{1,3}, 
Fengquan Wu\inst{1,3},Jean-Eric Campagne$^{2}$, Christophe Magneville$^{5}$ }
\institute{ Key Laboratory of Computational Astrophysics, National Astronomical Observatories, Chinese Academy of Sciences, Beijing 100012, China\\
  \and    Universit\'e Paris-Sud, LAL, UMR 8607, F-91898 Orsay Cedex, France $\&$ CNRS/IN2P3, F-91405 Orsay, France\\
  \and    University of Chinese Academy of Sciences, Beijing 100049, China\\
   \and    Center for High Energy Physics, Peking University, Beijing 100871, China\\
   \and CEA, DSM/IRFU, Centre d'Etudes de Saclay, F-91191 Gif-sur-Yvette, France
}
  
\abstract{ 
In this paper, we apply our sky map reconstruction method for transit type interferometers 
to the Tianlai cylinder array. The method is based on the spherical harmonic 
decomposition, and can be applied to cylindrical array as well as dish arrays and we can compute
the instrument response, synthesised  beam, transfer function and the noise power spectrum. 
We consider cylinder arrays with feed spacing larger than half wavelength, and as expected, we find
that the arrays with regular spacing have grating lobes which produce spurious images in the reconstructed 
maps. We show that this problem can be overcome, using arrays with different feed spacing on each cylinder. 
We present the reconstructed maps, and study the performance in terms of noise power spectrum, transfer function 
and beams for both regular and  irregular feed spacing configurations.
\keywords{Cosmology: observation, HI intensity mapping; Method: transit telescope; map making  }
}

\authorrunning{J. Zhang et al.}            
\maketitle


\section{Introduction}           


The determination of the neutral hydrogen (\HI) distribution from 21cm line observation is an important method 
to study the statistical properties of Large Scale Structures in the Universe.
The intensity mapping technic is an efficient and economical way to map the universe using the (\HI) 21cm emission,
which is suitable for late time cosmological studies (($z \lesssim 3$)), 
especially for constraining dark energy models through baryon acoustic oscillation (BAO) features
\citep{2006astro.ph..6104P,2008PhRvL.100i1303C,ansari.IM.2008,2012A&A...540A.129A,2010ApJ...721..164S,2011ApJ...740L..20G}. 
Large wide-field and wide band radio telescopes would be needed to achieve rapidly the observation of 
large volumes of the universe. Several dedicated experiments are aimed at such surveys, including our own experiment Tianlai\footnote{\tt http://tianlai.bao.ac.cn} \citep{2012IJMPS..12..256C}, as well as CHIME \citep{2014SPIE.9145E..22B}, BINGO \citep{2013MNRAS.434.1239B},
HIRAX\footnote{\tt http://www.acru.ukzn.ac.za/cosmosafari/wp-content/uploads/2014/08/Sievers.pdf}, and
BAORadio\footnote{\tt http://groups.lal.in2p3.fr/bao21cm}.

In transit mode intensity mapping surveys, the antennas are fixed on the ground during observation, it observes the 
sky as the Earth rotates. For the cylinder arrays such as Tianlai and CHIME, 
the instantaneous field of view is a strip of sky along the meridian, and sky patches of different right ascension pass through the field of view. 
As the telescopes do not need to track the celestial target, the mechanical structure of the telescope is very simple.

The Tianlai project is designed to survey the large scale structure by intensity mapping of the redshifted 21cm line, 
and to constrain dark energy models by baryon acoustic oscillation (BAO) measurement. As a first step, the current 
pathfinder experiment will test the basic principles and key technologies of the 21cm intensity mapping method.
The Tianlai array is a wide band interferometer which features both a cylinder array and a dish array,
installed at a radio quiet site ($44^\circ 10' 47'' N, 91^\circ 43' 36'' E$) in Hongliuxia, Balikun County, Xinjiang 
Autonomous Region in northwest China \citep{2015IAUGA..2252187C}. 
The construction of the Tianlai cylinder and dish pathfinder arrays have been completed at the end of 2015, 
the two arrays are now undergoing the commissioning process.  
The map making algorithm and its application to dish arrays has been presented 
in \citet{2016arXiv1606.03090Z}, heretofore referred to as {\bf paper I}.  In the present paper, we will focus on its application 
to the Tianlai pathfinder cylinder array.

The Tianlai cylinder pathfinder array has  three adjacent cylindrical reflectors oriented in the North-South direction, 
each cylinder is 15 m wide and 40 m long. At present the cylinders are equipped with a total of 96 dual polarization receivers
which do not cover the full length of the cylinders. In the future, the pathfiner instrument may be upgraded by simply adding 
more feed units and the associated electronics. 
The longer term plan is to expand the Tianlai array to full scale once the principle of intensity mapping is proven to work. 
The full scale Tianlai cylinder array would have a collecting area of $\sim (10^4 \, \mathrm{m})^2$, and 
$\sim 10^3$ receiver units. 
A forecast for its capability in measuring dark energy and constrain primordial non-Gaussianity can be 
found in \citet{2015ApJ...798...40X}. In addition to the redshifted 21 cm intensity mapping observation, such surveys  
may also be used for other observations, such as 21cm absorber\citep{2014PhRvL.113d1303Y}, fast radio burst (FRB) \citep{2015Natur.528..523M,2016arXiv160207292C}, and electromagnetic counter part of gravitational wave events \citep{2014arXiv1405.6219F}.

\begin{figure}
\centering
\includegraphics [width=0.5\textwidth] {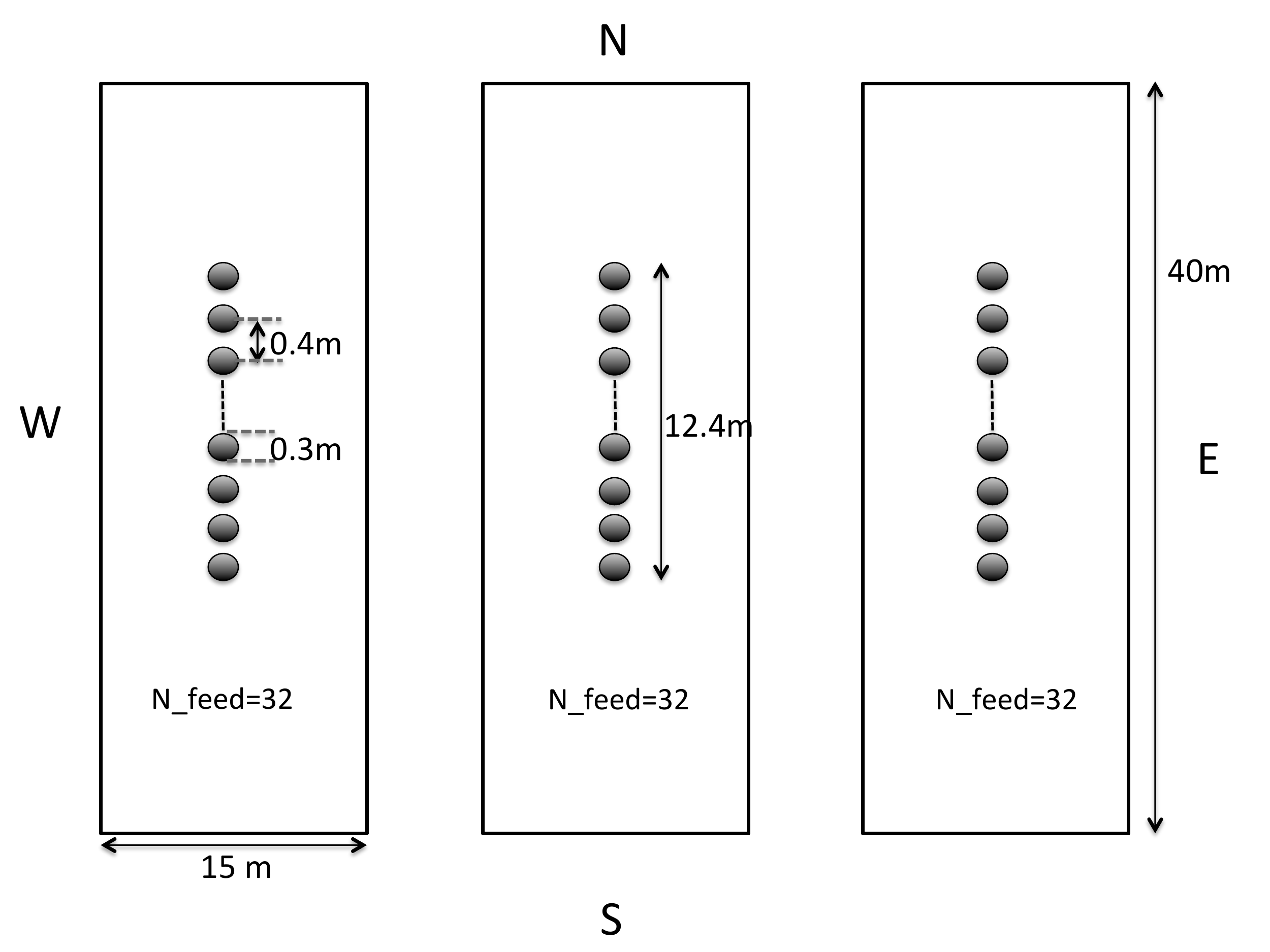}
\caption{Regular configuration of the cylinder array.}
\label{fig-config}
\end{figure}

The simplest arrangement of the existing 96 feeds is to have 32 feeds on each cylinder, regularly spaced
so that on each cylinder the feeds form a uniformly spaced linear array.  
Two such configurations are considered here:
\begin{itemize}
\item[Regular 1.]  The feed spacing is taken to be   $d_{\rm sep} = 0.4$m, which is 
about one wavelength at the observation frequency of 750 MHz. In this configuration, the feeds occupies only 12.4 m of 
the total 40 m length of the cylinder, as as shown on Fig\ref{fig-config}. 
\item[Regular 2.]  The feed spacing is taken to be  $d_{\rm sep} = 0.8$m, about twice the wavelength at the 
cylinder, 
\end{itemize}

One may also consider configurations with irregular positioning of the feeds to reduce grating lobes. In this paper we consider a very 
simple extension: on each cylinder the feeds still forms a uniform linear array, but the number of feeds and hence the spacing of the array 
is different on each cylinder. We have total 96 feeds at the present time. Marking the cylinders from East to West as 
cylinder1, cylinder2 and cylinder3 respectively, we consider the following configurations:
\begin{itemize}
\item[Irregular 1.]  This is the first irregular cylinder array with number of feeds on each cylinder 31, 32 and 33 respectively. 
The feeds occupy  12.4 m along NS direction on each cylinder.
The feed spacing would be $d_{\rm sep} = 0.413m$ for cylinder1, $d_{\rm sep} = 0.4$m for cylinder2 and  $d_{\rm sep} = 0.388$m for cylinder3.
\item[Irregular 2.]  This is the second irregular cylinder array with number of feeds on each cylinder 31, 32 and 33 respectively, but the feeds
occupy 24.8 m along NS direction on each cylinder.
The feed spacing would be $d_{\rm sep} = 0.827m$ for cylinder1, $d_{\rm sep} = 0.8$m for cylinder2 and 
$d_{\rm sep} = 0.775$m for cylinder3 for this 
\end{itemize}

\begin{figure}
\centering
\includegraphics [width=0.8\textwidth] {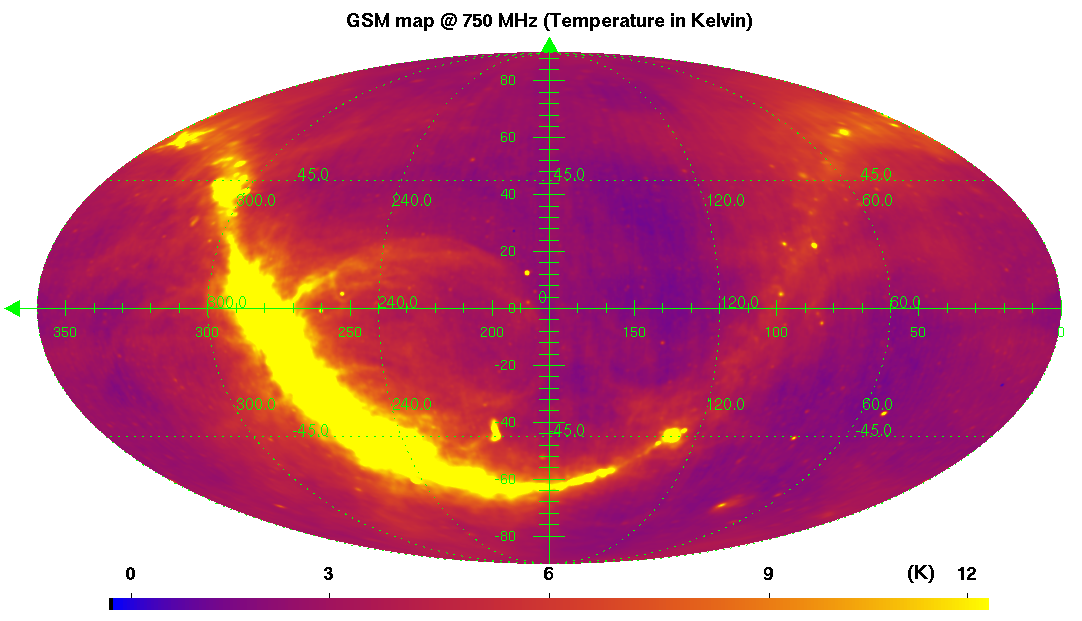}
\caption{GSM (Global Sky Model) map at 750 MHz, used as the true sky for the reconstruction with the Tianlai cylinder configuration.}
\label{fig-inmap}
\end{figure}

To simulate the map making process, we use an input map based on the Global Sky Model (GSM) \citep{2008MNRAS.388..247D},
shown in Fig. \ref{fig-inmap}.  The map is obviously dominated by the radiation from the galactic plane, which is mostly synchrotron emission from galactic cosmic ray electrons. 
For the computations carried out in this work, we have used  HEALPix  \citep{2005ApJ...622..759G} to pixellate the celestial sphere, 
with $n_{side} = 512$. In our spherical harmonics transformation we take $\ell_{max}=1500$, which is sufficient for the angular resolution 
of Tianlai pathfinder cylinder array. 

Below, we present in Sec. \ref{sec-method} a brief review of the spherical harmonic decomposition map making method. In 
Sec. \ref{sec-regular} we discuss the grating lobe problem and spurious image for regular receivers layout.  
To resolve this problem, in Sec. \ref{sec-irregular} we study the case of irregular layouts listed above.  
We present our conclusion in section \ref{sec-conclusion}.

\section{A brief review of sky reconstruction method}
\label{sec-method}
In this section, we present briefly the map making method for transit interferometer array  based on spherical harmonic decomposition.
A more detailed presentation and comparison with the classical radio interferometry (tracking type surveys) can be found in paper I, 
as well as in \citet{2014ApJ...781...57S}.
Unlike the frequently used tracking observations, 
it is more convenient to work in ground coordinates in which the baselines of the array do not change in the transit observation. In this 
formalism, the visibilities recorded as a function of time correspond to observations of different parts of the sky.
We separate the inversion problem into independent sub-systems using m-mode decomposition in spherical harmonics
and we assume that the individual feed responses and array geometry are known.
The sky emission intensity is $I(\vec{\hat{n}}) = \langle \vec{E}^*(\vec{\hat{n}},t) \cdot \vec{E}(\vec{\hat{n}},t) \rangle_t $,
where $\vec{\hat{n}}=(\alpha, \delta)$ denotes the sky direction, and the
receivers are sensitive to the sky emission complex amplitudes $\vec{E}(\vec{\hat{n}})$.
A single receiving element can be characterized by its angular complex response $D(\vec{\hat{n}})$ and its position 
$\vec{r}$, the output of 
element $j$ is 
\begin{equation}
s_j (t) =  \iint d \vec{\hat{n}}  \, D_j(\vec{\hat{n}},t) \,E(\vec{\hat{n}},t)\, e^{i \vec{k} \cdot \vec{r_j} }  
\end{equation}
The Visibility $V_{ij}=  <s_i^* ~ s_j >_t$ is the short time average of the cross correlation of the 
outputs of a pair of antennae or feeds $s_i, s_j$, located at positions $\vec{r_i}, \vec{r_j}$ with $\vec{r_{ij}} \equiv \vec{r_j} - \vec{r_i}$. 
As the emission of the different sky directions are incoherent, only the wave from the same sky direction are correlated, and the integration 
yields the interferometer equation
\begin{eqnarray}
V_{ij}   & = & \iint \, D_i^*(\vec{\hat{n}}) \, D_j(\vec{\hat{n}}) \,I(\vec{\hat{n}}) \,  e^{i \vec{k} \cdot \vec{b_{ij}} \cdot } \, d \vec{\hat{n}} \\
&=& \iint \, L_{ij}(\vec{\hat{n}})\,  I(\vec{\hat{n}}) \, d \vec{\hat{n}},
\label{eq-visib}
\end{eqnarray} 
In a transit instrument, visibilities are measured as a function of time or right ascension, 
with the beam response $ L_{ij}(\vec{\hat{n}})$ changing due to earth rotation. In discrete form, including the noise contribution 
and gathering visibility measurement from all baselines and from all times into a vector, we can write the full 
measurement equation in matrix form: 
\begin{equation}
\label{eq-visI}
[\mathrm{V}] = [\mathbf{L}] [\mathrm{I}] + [n].
\end{equation}
where we have used the square brackets to emphasis these are vectors and matrices, and the time ordered visibility data 
$[\mathrm{V}] $ is linearly related to the sky intensity of different directions $\mathrm{I} $ by the time dependent 
beam response represented as the matrix $[ \mathbf{L} ]$. 
The interferometer array map-making is to solve this system and reconstruct $\mathrm{I} $ from the observed time ordered 
visibility data.

We expand the sky intensity and beam response into spherical harmonics, 
\begin{eqnarray}
\label{eq-mapdecomp}
I(\vec{\hat{n}}) & = & \sum_{\ell = 0}^{+\infty}  \sum_{m=-\ell}^{+\ell} \, \mathcal{I}_{\ell , m} \, Y_{\ell,m} (\vec{\hat{n}}), \\ 
L_{ij}(\vec{\hat{n}}) & = & \sum_{\ell = 0}^{+\infty}  \sum_{m=-\ell}^{+\ell} \, \mathcal{L}_{\ell , m} \, Y_{\ell,m} (\vec{\hat{n}}).
\label{eq-beampattern}
\end{eqnarray} 
where $(\theta,\phi)$ are the polar coordinates, $\theta=90^\circ - \delta$, $\phi =\alpha$. $Y_{\ell,m}(\vec{\hat{n}})$ are the spherical harmonic functions. Then the visibilities can also be written as a summation of spherical harmonic modes,
$$V_{ij} =   \sum_{\ell,m}  (-1)^m \mathcal{I}_{\ell , m} \, \mathcal{L}_{\ell , -m}$$

For the transit interferometer array, the effect of the Earth rotation is that the beam $L_{ij}(\vec{\hat{n}})$ has a 
constant drift along the right ascension direction, with the offset angle given by $\alpha_p(t)=\alpha_{0}+ \Omega_e \, t$,  
where $\Omega_e$ is angular rotation rate of the Earth, so that
\begin{equation}
L_{ij}(\vec{\hat{n}},t) = L_{ij}((\theta, \varphi),t) =  L_{ij}(\theta, \varphi - \alpha_p(t)).
\end{equation}
The spherical harmonics coefficients of the rotated/shifted beams can be written as
$\mathcal{L}_{\ell , m} (t_k)  =   \mathcal{L}_{\ell , m} e^{- i m \alpha_p(t)}$.
The recorded visibilities as a function of  $\alpha_p$ are then
\begin{eqnarray}
V_{ij} (\alpha_p)  & = &  \sum_{m=-\infty}^{+\infty} \sum_{\ell = |m| }^{+\infty}  \, (-1)^m \, 
\mathcal{I}_{\ell , m}  \, \mathcal{L}_{\ell , -m}   \, e^{i m \alpha_p} 
\end{eqnarray}
We recognise the expression as a Fourier series for the periodic function $ V_{ij} (\alpha_p) $; the corresponding 
Fourier coefficients $\tilde{\mathcal{V}}_{ij}(m)$, computed from a set of regularly time sampled visibility measurements is
\begin{eqnarray}
\tilde{V}_{ij}(m)  =  (-1)^m \, \sum_{\ell = |m| }^{+\ell_{max} } \, \, \mathcal{I}_{\ell , m}  \, \mathcal{L}_{\ell , -m} + \mathrm{noise}
\label{eq-vis-I-FFT}
\end{eqnarray}
Grouping m-mode visibilities from all baselines in a vector and using matrix notation, we can write the measurement 
equation for each m-mode as: 
\begin{eqnarray}
\left[  \tilde{V}_{ij}  \right]_m & = &   \mathbf{L}_{ij,m} \,  \times \, \left[  \mathcal{I} (\ell) \right]_m +  \left[  \tilde{n}_{ij}  \right]_m
\label{eq-visibility_ij}
\end{eqnarray}
or putting all baselines of the array together, 
\begin{eqnarray}
\left[  \tilde{V}   \right]_m & = &   \mathbf{L}_{m} \,  \times \, \left[  \mathcal{I} (\ell) \right]_m +  \left[  \tilde{\mathbf{n}}  \right]_m
\label{eq-visibility}
\end{eqnarray}
Comparing with Eq.(\ref{eq-visI}),  the full linear system is decomposed into a set of $m_{max}=\ell_{max}$ 
independent smaller system, one for each $m$-mode, which have much smaller dimensions ($n_{\rm baseline} \times \ell_{max})$
and are thus much easier to solve numerically.
 
We assume that the noise on visibility measurement follows a gaussian random process, with variance
$\mathbf{N}=<n n^\dagger>$. Using maximum likelihood method,  the solution of observed sky is given by
\begin{equation}
[ \mathcal{\widehat{I}} ]_m  = \mathbf{L}_m^{-1} \, [ \tilde{V} ]_m 
\label{eq-synthesis}
\end{equation}
where $\mathbf{L}_m^{-1}$ denotes the noise weighted pseudo-inverse matrix of $\mathbf{L}_m$. 
This can be computed by using the singular-value-decomposition (SVD) method: any $m\times n$ matrix $\mathbf{A}$ can be decomposed 
as $\mathbf{A}=\mathbf{U \Sigma Q^{\dagger}}$,
where $\mathbf{U}$ and $\mathbf{Q}$ are $m \times m$ and $n \times n$ unitary matrices, and $\mathbf{\Sigma}$ 
is an $m \times n$ rectangular diagonal matrix,  i.e. all non-diagonal elements are zero, with non-negative real numbers
on the diagonal. The pseudo-inverse is given by 
\begin{equation}
\tilde{\mathbf{A}}^{-1} = \mathbf{Q} \tilde{\mathbf{\Sigma}}^{-1} \mathbf{U}^\dagger
\end{equation}
where $\tilde{\mathbf{\Sigma}}^{-1}$ is obtained by replacing all diagonal elements $e_{ii}$ above certain threshold value by their
reciprocal $1/e_{ii}$, while setting the other elements zero. For details of computing the pseudo-inverse, see e.g. \citet{2012BrJPh..42..146B}.

Substitute Eq.~(\ref{eq-visibility}) to Eq.~(\ref{eq-synthesis}),  and neglecting the noise, we have 
$\mathcal{\hat{I}}= \mathcal{R} \mathcal{I}$, where $\mathcal{R}$ denotes the reconstruction or response matrix, 
which relates the reconstructed sky to the original sky. In the spherical harmonics representation, 
the $m$-mode reconstruction matrix is $\mathbf{R}_m \equiv \tilde{\mathbf{L}}_m^{-1} \mathbf{L}_m$.
Ideally, if $\mathbf{R}_m=\mathbf{I}$ then the reconstruction for the $m$-mode is completely accurate. In practice, 
the reconstruction is usually not fully accurate.
 
For each given $m$, the different $\ell$ coefficients are correlated, the physical measurement data is a mix of 
different $\ell$ mode contributions. 
We can define the compressed response matrix $\mathbf{R}$ by extracting the diagonal terms from individual $\mathbf{R}_m$ matrices:
\begin{eqnarray*}
\mathbf{R}(\ell, m) & = & \mathbf{R}_m (\ell , \ell) 
\label{eq-Rmatrix} 
\end{eqnarray*}
Obviously, the $\mathbf{R}(\ell,m)$ do not describe fully the reconstruction in the $(\ell,m)$ plane and 
the original $\mathbf{R}_m$ matrices are needed. However, the $\mathbf{R}(\ell,m)$ matrix can give some idea 
of how well an $(\ell, m)$ mode is measured with a given array configuration, 
so it can help us to compare the performance of different configurations.

If we consider the reconstruction of sky spherical harmonics coefficients from pure noise visibilities
$(\tilde{\mathrm{V}}_{ij} = \tilde{n}_{ij} )$, the covariance 
matrix $\mathbf{Cov} (\ell_1 , \ell_2)$ of the estimator $\widehat{\mathcal{I}}(\ell,m)$ for each mode $m$ 
can be computed from the $\mathbf{L}_m^{-1}$ matrix and the noise covariance matrix 
$\mathbf{N}_m =  \left[ \tilde{\mathrm{n}}_{ij} \right]_m \cdot \left[ \tilde{\mathrm{n}}_{ij} \right]_m^\dag$, 
\begin{eqnarray*}
\mathbf{Cov}_m (\ell_1 , \ell_2) & = & \langle \left[ \widehat{\mathcal{I}}(\ell) \right]_m \cdot \left[ \widehat{\mathcal{I}}(\ell) \right]_m ^\dag \rangle  = \mathbf{L}_m^{-1} \, \mathbf{N}_m \,  \mathbf{L}_m^{-1 \dag}    
\end{eqnarray*}
The covariance matrix is not diagonal, especially due to partial sky coverage in declination. 
However, if we ignore this correlation and use the diagonal terms only for each $m$ mode, we can gather them 
together to create the $ \sigma_{\mathcal{I}}^2(\ell,m)$ variance matrix. This matrix informs us on how well each 
$(\ell,m)$ mode is measured. 
\begin{eqnarray}
\sigma_{\mathcal{I}}^2(\ell,m)  & = & \mathbf{Cov}_m (\ell , \ell)  \label{eq-covardiag} 
\end{eqnarray}

We consider a survey duration of two full years for the results presented in this paper.
The total integration time for each visibility time sample would be $t_{int} \sim 2\times10^4 \, \mathrm{s}$ 
for $n_t=2 m_{max}=3000$. Assuming a system temperature 
$T_{sys} = 50 \, \mathrm{K}$, and $\Delta \nu=1 \, \mathrm{MHz}$, 
the effective $\sigma_{noise}$ for measured visibility time samples can then be 
written as a function of integration time per time sample $t_{int}$:
\begin{eqnarray}
\sigma_{noise} & = & \frac{\sqrt{2} T_{sys}}{ \sqrt{t_{int} \, \Delta \nu} } \sim  0.49 \, \mathrm{mK} \label{eq-signoise-1}  
\end{eqnarray}

\section{The regular array configuration}
\label{sec-regular}

\begin{figure}[htbp]
\centering
\includegraphics [width=0.75\textwidth] {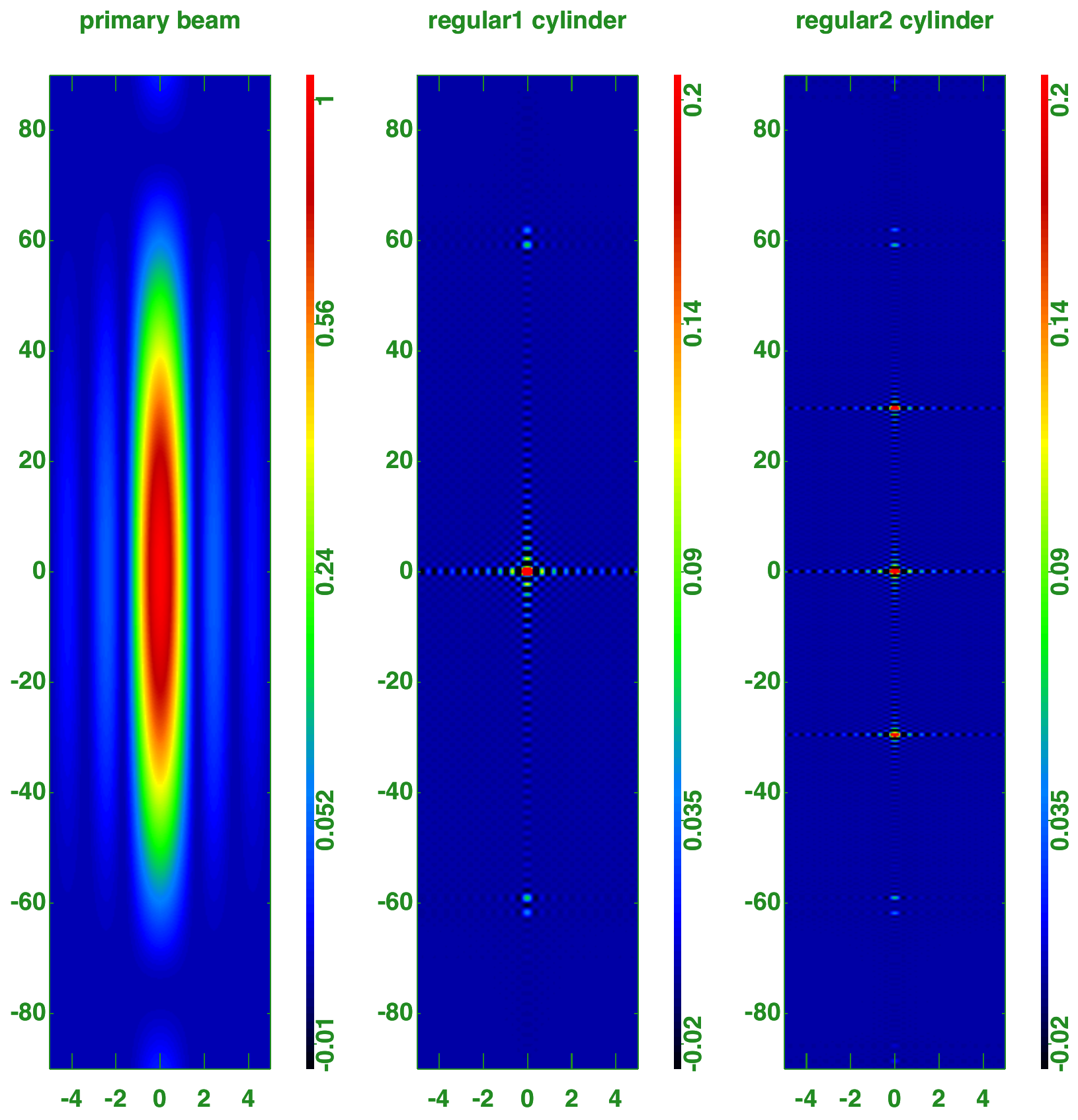}
\caption{The primary beam (left) and synthetic beams for the regular 1 (center) and regular 2 (right) configurations. }
\label{fig-beam-re}
\end{figure}

The primary beams for each feed on the cylinders is narrow in the East-West (EW) direction and wide in the North-South (NS) direction, as 
determined by the cylinder reflector curvatures. We model the primary beam of a single feed on associated with a cylindrical reflector as 
\begin{equation}
D(\alpha, \beta) \propto \frac{\sin ( \alpha \pi (L_x / \lambda)  ) }{ \alpha \pi (L_x / \lambda)} \frac{\sin ( \beta \pi (L_y / \lambda) ) }{ \beta \pi (L_y / \lambda)}   
\label{eq-primarybeam}
\end{equation}
where $(\alpha, \beta)$ are the two angles with respect to the feed axis, along the EW and NS planes respectively. 
$\lambda$ is the wavelength,  $L_x $ and  $L_y $ are the effective feed sizes along the EW and NS planes.
We take $L_y =0.3$m for the Tianlai cylinder feeds, and $L_x=13.5$m  corresponding to an illumination efficiency of 
0.9 for the feed on the 15m wide cylinder. These parameters gives a beam width $\sim 100^\circ$  
in the North-South direction, and $\sim 2^\circ$ in the East-West direction at 750 MHz, 
The actual values will be obtained by fitting the real observational data, these are heuristic 
values  but should be sufficient for our estimations here. 
The primary beam is shown as the left panel of Fig. \ref{fig-beam-re}.

\begin{figure}
\centering
\includegraphics [width=0.9\textwidth] {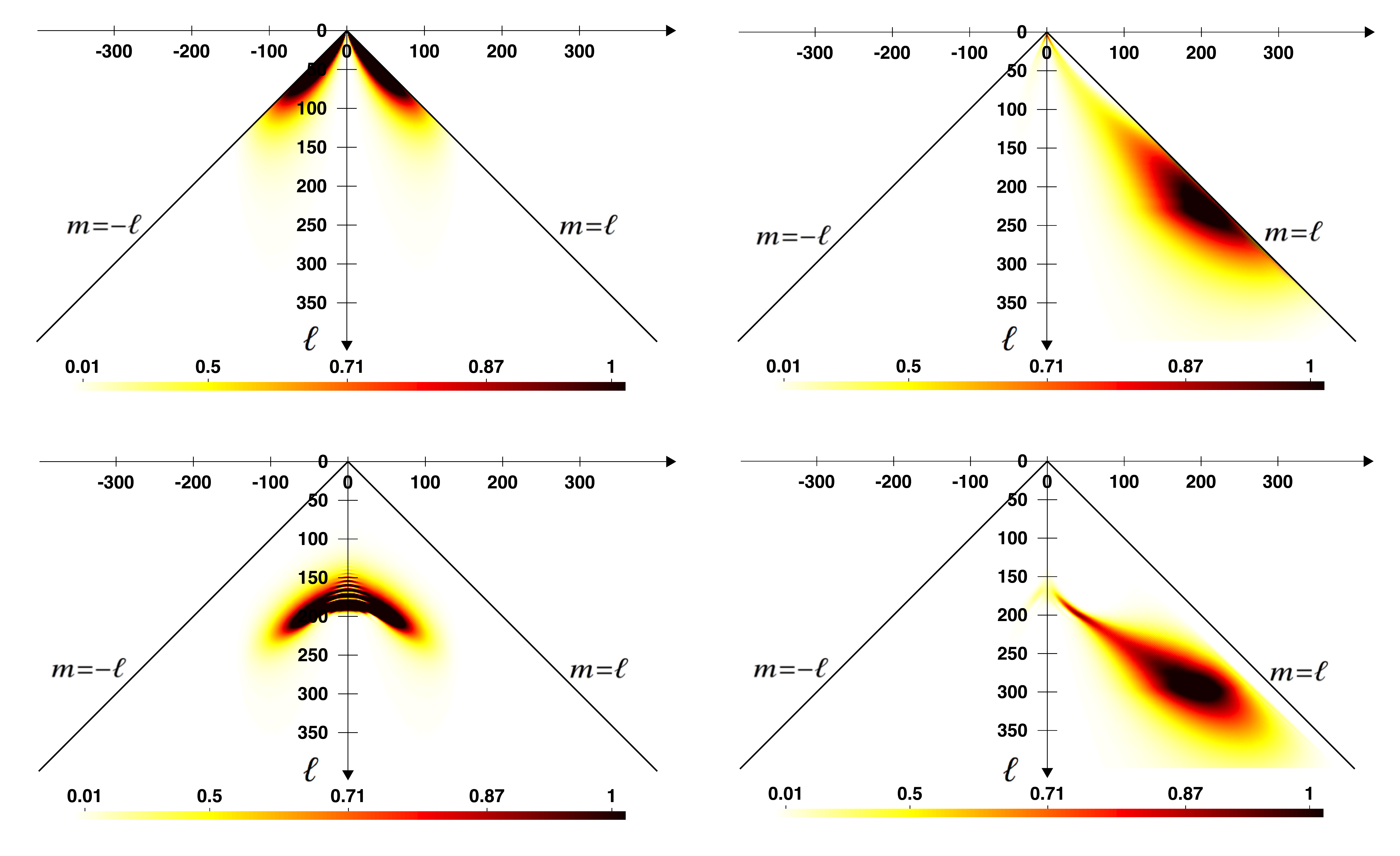}
\caption{The beam patterns in spherical harmonics $\mathcal{L}_{\ell,m}$ with size $L_x = 13.5$m, $L_y =0.3$m and centered at
latitude $44.15^\circ$. Top Left: auto-correlation of a feed; Top Right: cross-correlation for a EW baseline with $d_{\rm sep}=15$m; Bottom Left: cross-correlation beam for a NS baseline with $d_{\rm sep}=12$m; 
Bottom Right: cross-correlation for a SE-NW baseline with $(\Delta x, \Delta y)=$(15m, 12m). }
\label{fig-beam-alm}
\end{figure}

For uniformly spaced linear arrays, grating lobes appear when the spacing is larger than half wavelength ($d_{\rm sep}>\lambda/2$). 
This is because the phase factor $\exp(i 2\pi d_{\rm sep} \sin \theta/\lambda )$ is periodic with respect to $\sin\theta $, and when 
$ d_{\rm sep}>\lambda/2$ the maximum appears more than once. We show the synthetic beam for the regular case 1 
and regular case 2 in the central and right panels respectively in Fig. \ref{fig-beam-re}. These are obtained by making the 
full synthesis of a point source image located at the latitude of the array, i.e. $44^\circ 10'$. As we can see in the figure, 
there are strong grating lobes along the NS direction in the synthetic beams. The position of the $n$th order grating 
lobe is $\sim n \lambda/d_{\rm sep}$. At 750MHz,  the positions are $\pm 57.3^\circ$ for the regular case 1 ($d_{\rm sep} = 0.4$m)
 and $\pm 28.6^\circ$ for the regular case 2 ($d_{\rm sep}=0.8$m).  In addition, there are also primary beam side lobes along both the NS and EW direction, these are less prominent in their level and has smaller periods.

To have a better understanding of the synthetic beams in the spherical harmonic space, 
let us consider the beams of a single pair of receivers.
In Fig.\ref{fig-beam-alm} we show the beam patterns function for four cases: the auto-correlation (top left), and
the cross-correlations for a due EW baseline between two cylinders (top right), for a due NS baseline (bottom left), and 
a SE-NW baseline (bottom right). By definition, only the region $-\ell<m<\ell$ has valid values. 
In the dish case (see paper I), the autocorrelation covers a triangular region with the top at the origin
 $(\ell,m)=(0,0)$, two sides and extends along $m = \pm \ell cos \delta$ where $\delta$ is declination of the observation, 
 and up to $\ell_{max}=2\pi D/\lambda$ where $D$ is the effective aperture. The auto-correlation in the cylinder case is very different, 
 taking up a butterfly shape. This is because the cylinder primary beam is asymmetric in the NS and EW direction.  As described in 
 Eq.~(\ref{eq-primarybeam}),  along the NS direction which corresponds to $m \sim 0$ the primary beam has very low resolution, 
 while along the EW direction the cylinder primary beam is about $\sim 2^\circ$ at 750 MHz, which corresponds to 
 $\ell \sim 2\pi L_x/\lambda \sim 210$. Indeed, the figure shows that the auto-correlation function extends
substantially to $\ell \sim 210$ along the two wings. Also, since the cylinder has almost the whole observable sky in its field of view, 
which includes the equator, the $m=\pm \ell$ is saturated. 

For the cross-correlations, the beam pattern centers at $(\ell,m) \sim (2\pi |\vec{u}|, 2\pi u)$
as expected, where $\vec{u} \equiv (u,v,w)=(b_x,b_y,b_z)/\lambda$. So the EW baseline is centered near $m \sim \ell$, while 
NS centered near $m=0$, with $\ell \sim 2\pi b/\lambda$. Note that here we are plotting only positive part of the baseline in one direction, 
so for the EW baseline the beam is on the $m>0$ side. If we are to plot the reverse direction, it would appear on the symmetric 
position at $m<0$.

\begin{figure}
\centering
\includegraphics [width=0.7\textwidth] {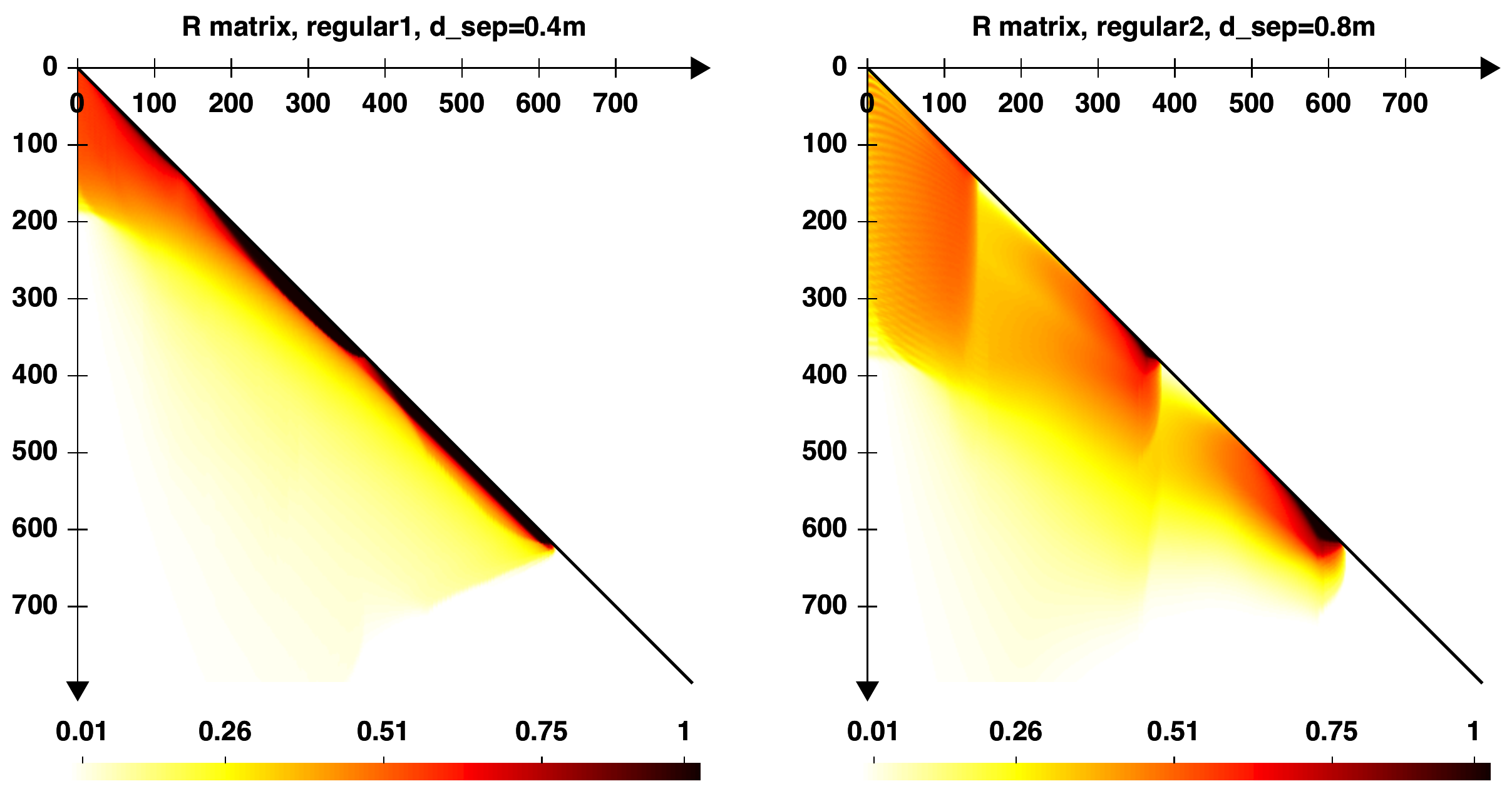}
\caption{Comparison of the R matrix for regular 1(left) and regular2 (right) configurations.}
\label{fig-R-reg}
\end{figure}
\begin{figure}
\centering
\includegraphics [width=0.7\textwidth] {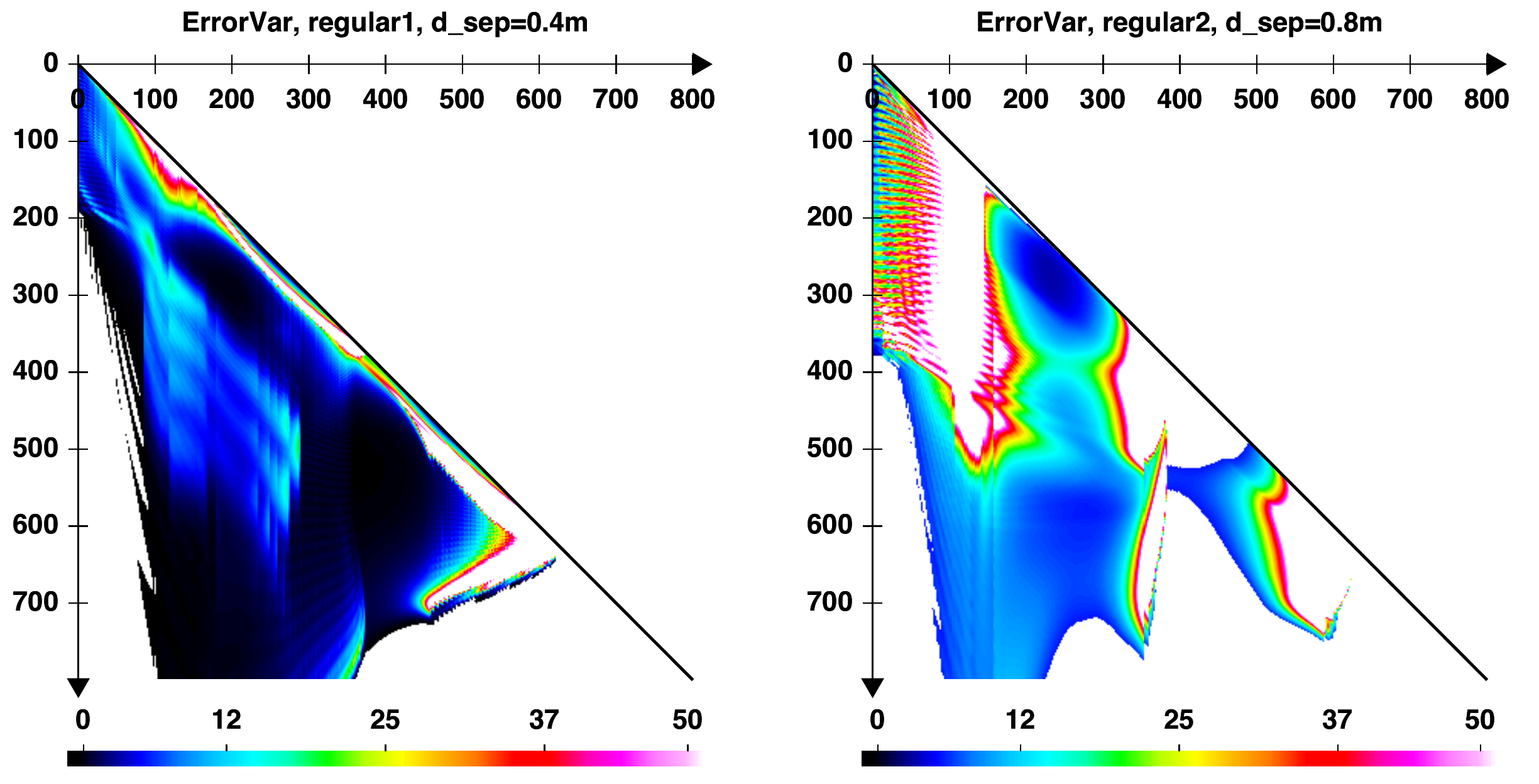}
\caption{Comparison of the error variance matrix for regular 1(left) and irregular2 (right) configurations. }
\label{fig-Errvar-reg}
\end{figure}

Figure \ref{fig-R-reg} shows the response matrix $\mathbf{R}(\ell,m)$ for the two regular configurations at frequency 750MHz.
In paper I, we noted that for each baseline the $\mathbf{R}$ matrix has a certain distribution centered at $(\ell,m)=2\pi b/\lambda$, 
where  $b$ is the baseline length. The $m$ position depends on both the EW component of the baseline and the declination of the 
strip to be observed. For an array with many baselines the $\mathbf{R}$ matrix are well described by the 
superposition of these individual baselines. For the cylinder array, the field of view is not limited to a narrow strip, but a
hemisphere or even larger spherical zone.  As such, the cylinder baseline would only be bounded by $m=\ell$. 
In the cylinder case, the $\mathbf{R}$ matrix at $m=0$ is significant up to $\ell \sim 190(380)$ for the Regular 1 (Regular 2)
case, which corresponds to the modes probed by the maximum baseline along one cylinder. The longest baseline of the array is 
however the diagonal ones, i.e. the baselines from North/South end of the East cylinder to the South/North end of the West cylinder,
so the $\mathbf{R}$ matrix distributes mostly on a band along $m=\ell$, with some extension to higher $\ell$ 
in the region between $m=0$ and $m=\ell$, forming a shark fin shape. The region near $m \sim 0$ is limited to relatively small $\ell$
due to the fact that our NS baselines are shorter, especially for the Regular 1 case. Future extensions which fill up 
the remaining part of the cylinder would help improve the $m=0$ region.  In the Regular 2 case, larger part of the $(\ell, m)$ 
space are covered than the Regular 1 case, but here the receivers are spread more widely, reducing the density of the baseline 
coverage, so here there are more apparent non-uniformity, as shown by the vertical stripes at $m=120$ and 350. These can be understood
as follows: as shown in Fig.\ref{fig-beam-alm}, each baseline is sensitive to some part of the $(\ell, m)$ space. The part of $(\ell,m)$
space which is not covered by baselines in the array would not be well reconstructed. As the cylinder array are aligned 
along the three cylinders, we can expect that the $m$ value centered at 0, 235, 470 will be covered, while regions
 between these, centered at $m=120$ and 350 are not well covered and may have large errors. Furthermore, if looking carefully, 
 some fringes near $m=0$ can also be seen, which may be due to the grating lobes.
 
 \begin{figure}[htbp]
\centering
\includegraphics [width=0.65\textwidth] {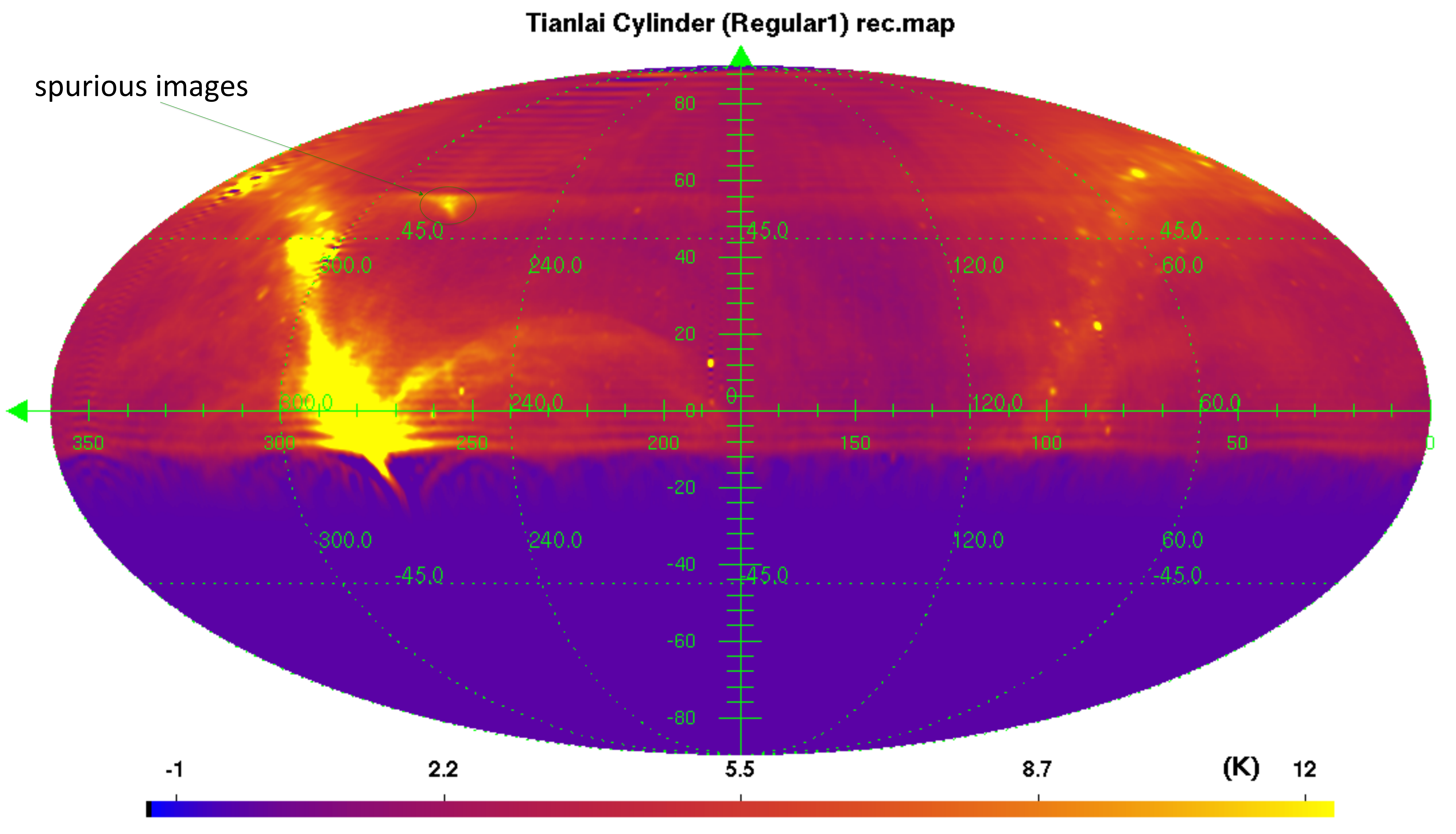}
\includegraphics [width=0.65\textwidth] {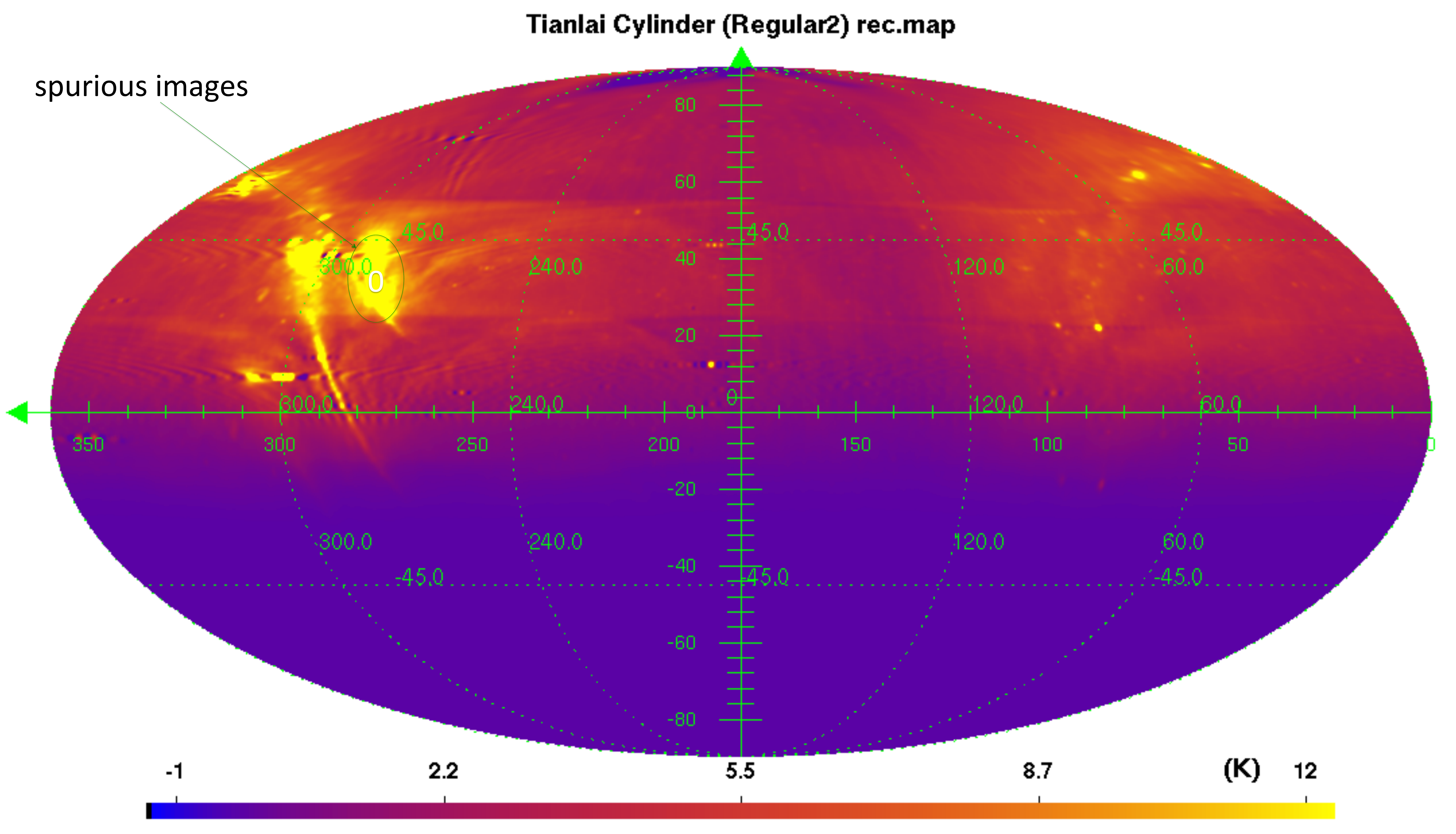}
\caption{Reconstructed sky map for the Tianlai cylinder configuration at 750 MHz. Top: Regular 2 configuration; Bottom: Regular 1 configuration. The input map is the GSM map at 750 MHz.}
\label{fig-recmap-regular}
\end{figure}

Figure \ref{fig-Errvar-reg} shows the corresponding error covariance matrix in the $(\ell, m)$ basis at 750 MHz. For the 
Regular 2 case the error is particularly large, but even for the Regular 1 case, the errors are also relatively large 
at these $m$ values.   The error values at other regions are relatively small. 
Additionally, in the Regular 2 case, near $m=0$ there is relatively large error and also the error shows some 
rapid modulation in $\ell$. These fringes are similar to the ones appeared in the $\mathbf{R}$ matrix at the same positions, 
and are due to the strong grating lobes.

In Fig.\ref{fig-recmap-regular} we show the reconstructed map at 750MHz derived from simulated regular cylinder array observation 
ignoring the instrument noise. The top figure use the Regular 1 configuration, and the bottom figure use the Regular 2 configuration.
Comparing with the original map Fig. \ref{fig-inmap}, there are spurious features appearing in the reconstructed map. This is 
very obvious for the Regular 2 case, and also present in the Regular 1 case (e.g. the bright spot at $(270^\circ, 54^\circ)$).  
These are produced  by the grating lobes of the brighter sources such as the Galactic plane and strong point sources, and the
Regular 2 case is worse than the Regular case 1. Because of such spurious features, one can not use array with such configurations 
to make reliable sky survey.

\section{The irregular array configuration}
\label{sec-irregular}

As we saw in the last section, spurious images appeared in the reconstructed maps of the regular array due to 
the presence of grating lobes. To avoid this problem, one could adopt spacings less than half wavelength, or 
employ non-uniform spacing in the linear array. However, at the wavelength of our observation, it is not practical to 
have spacing less than half wavelength. There are many possible non-uniform spacing schemes, 
here we choose a very simple one: adopting slightly different spacings on the three different cylinders. 
So we take the same total length on the three cylinders, 
but place 31, 32, and 33 feeds on each cylinder, so that the unit separations are different in each case. We choose the same
two total lengths as the regular cases described in last section. So for  the Irregular 1 case, 
the basic spacings are $d_{\rm sep}=0.413, 0.4, 0.388$m for the three cylinders respectively, with a total length of 12.4m; 
for the Irregular 2 case, the basic spacings are 0.827, 0.8 and 0.775 m respectively, with a total length of 24.8m.

\begin{figure}
\centering
\includegraphics [width=0.5\textwidth] {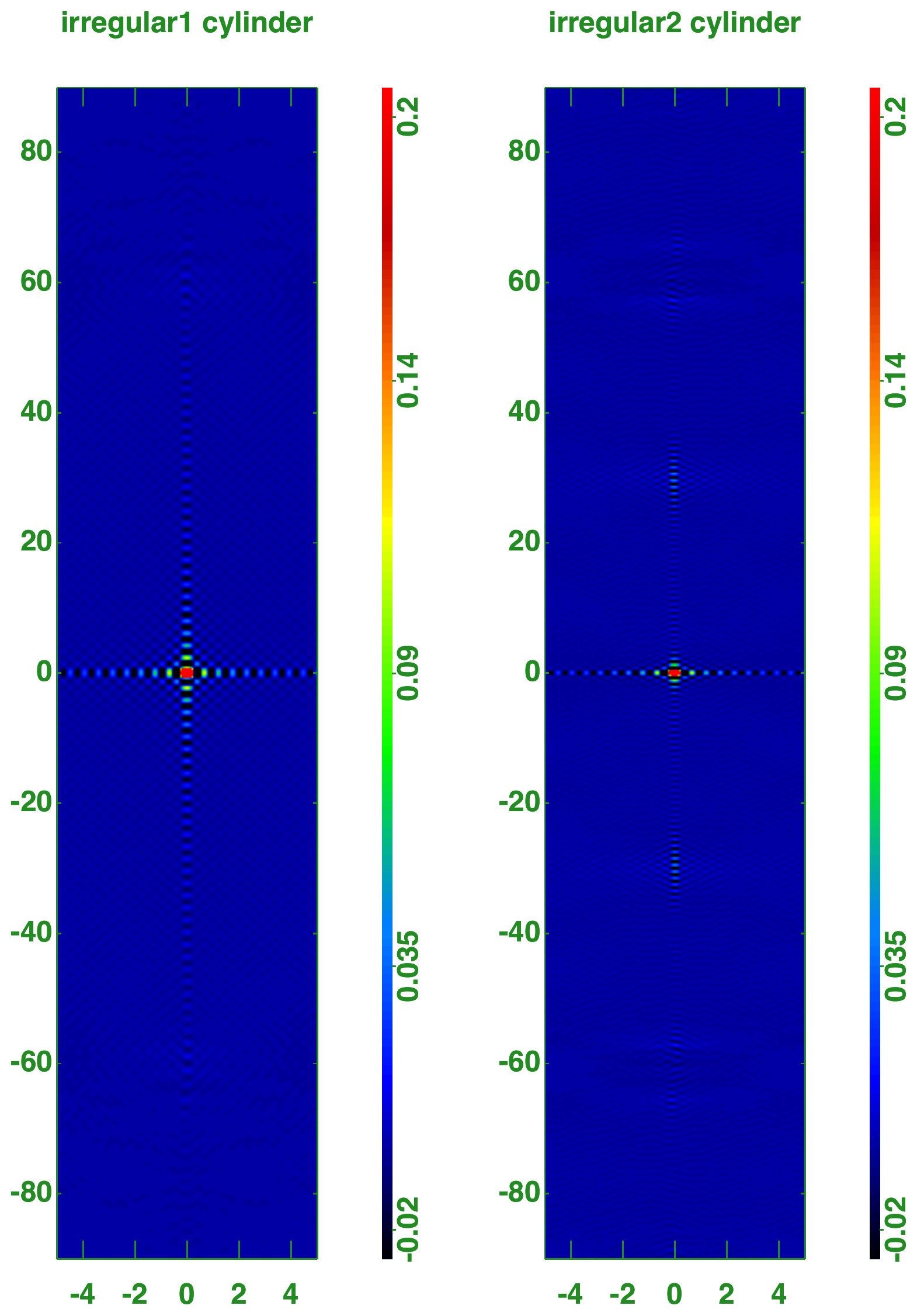}
\caption{The synthetic beam for Tianlai cylinder irregular configuration.}
\label{fig-beam-irre}
\end{figure}

There are still some degeneracies in the north-south baseline. For instance, there are 30, 31, 32 instances of 
$d_{\rm sep}=0.413, 0.4, 0.388$m NS baselines in the Irregular 1 configuration, respectively. Nevertheless, for the whole array 
there are NS baselines of different lengths. The slightly different positioning of the 
receivers also creates baselines  which deviates from the EW direction to different degrees. The whole set up allows
wider and more uniform coverage on the $(\ell, m)$-plane. 

Fig. \ref{fig-beam-irre} shows the synthesized beam for the two irregular cases. 
Here we see that the level of grating lobes is greatly reduced. Whereas in Fig.\ref{fig-beam-re} we can see clearly the 
sharp grating lobes at  $28^\circ$ for Regular 2 configuration and at $57^\circ$  for both Regular 1 and Regular 2 configurations,
in Fig. \ref{fig-beam-irre} at these angles the lobes are barely visible.  Of course, there are still the primary beam side lobes, but 
these are generally much smaller. Here we note that the Irregular 2 lobes are weaker than the Irregular 1 lobes.

\begin{figure}
\centering
\includegraphics [width=0.7\textwidth] {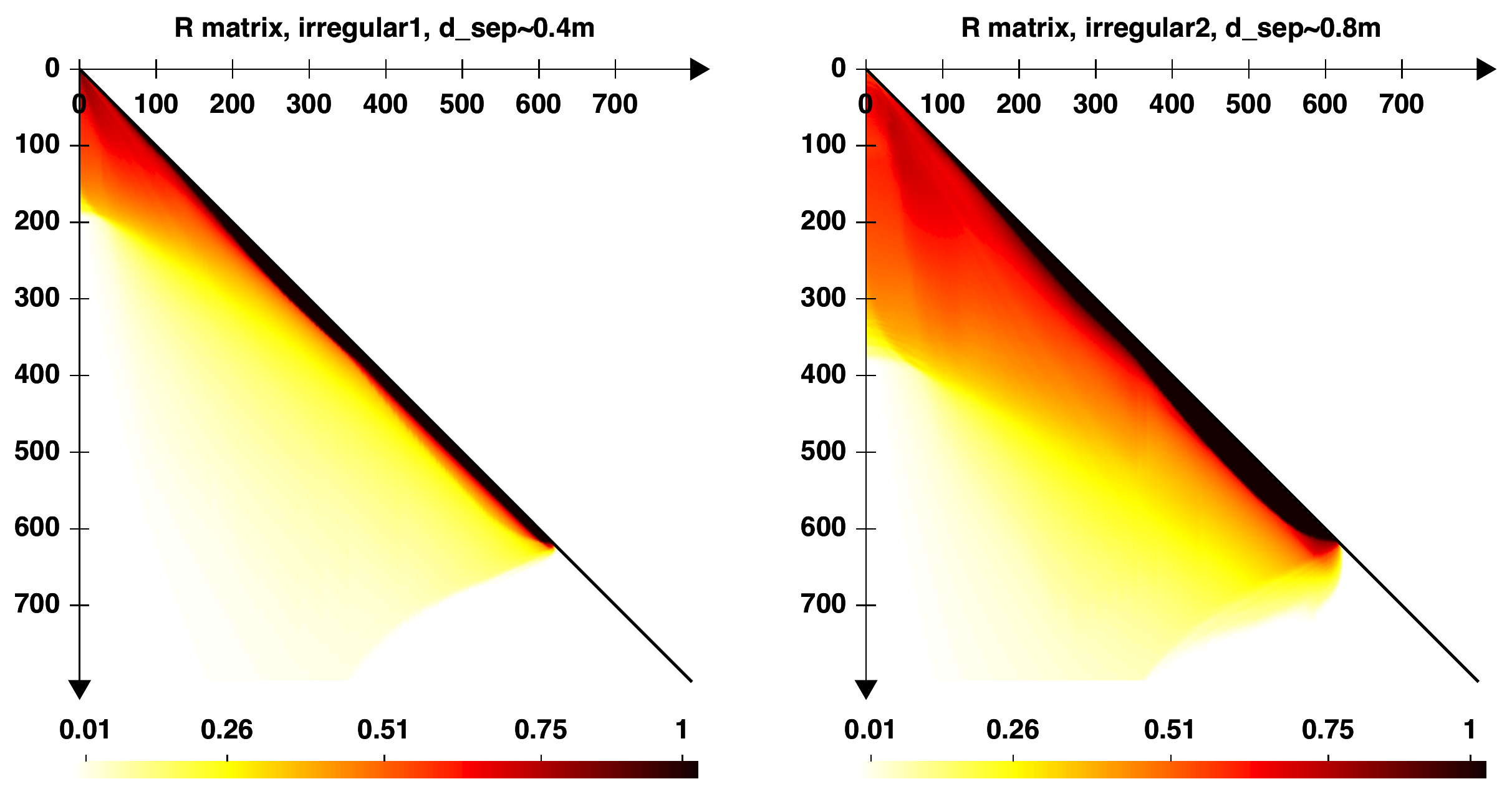}
\caption{Comparison of the R matrix for the Irregular 1 (left) and Irregular2 (right) configurations.}
\label{fig-R-irre}
\end{figure}
\begin{figure}
\centering
\includegraphics [width=0.7\textwidth] {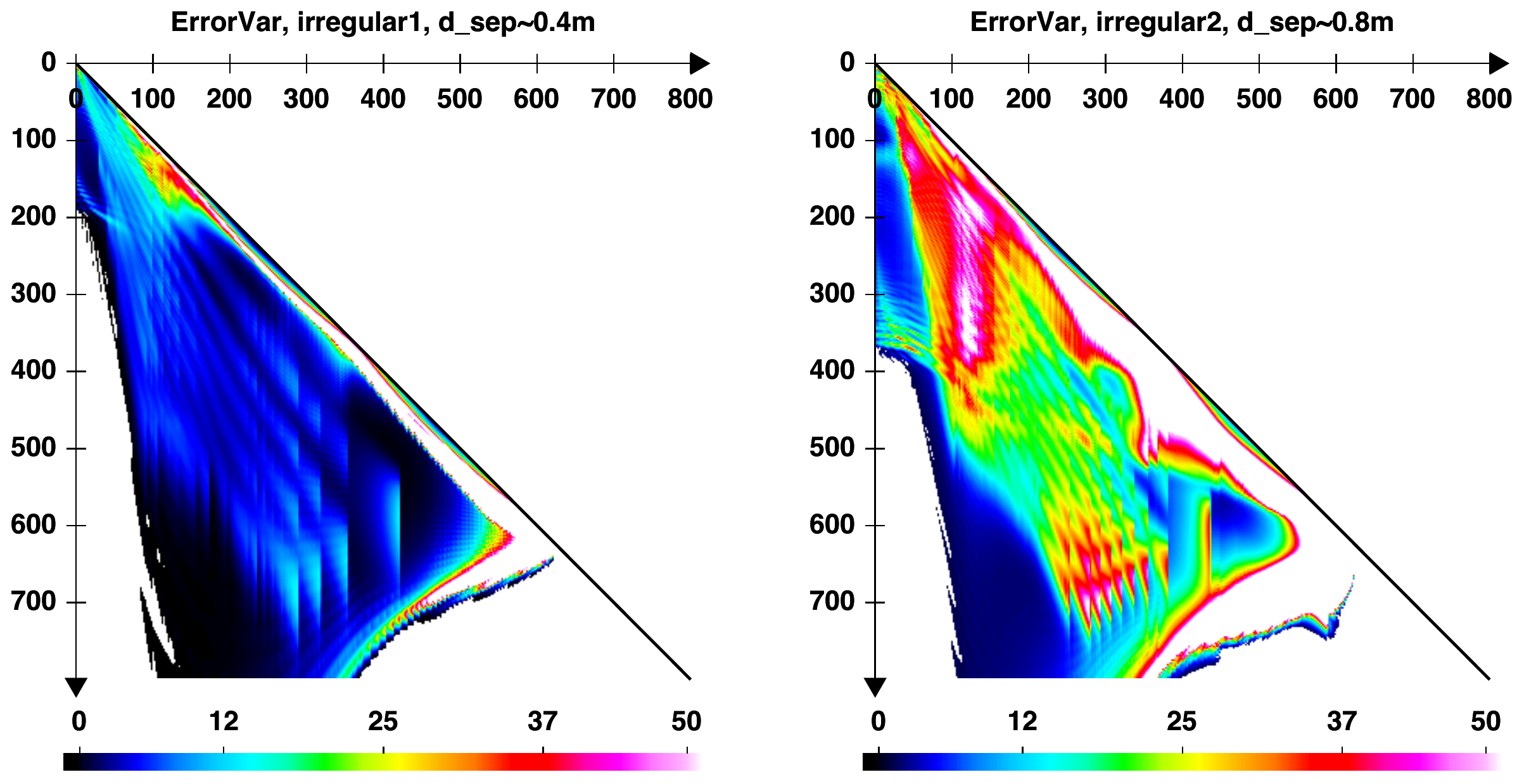}
\caption{Comparison of the error variance matrix for irregular configurations. 
}
\label{fig-Errvar-irre}
\end{figure}

We plot in Figure \ref{fig-R-irre} the compressed response matrix $\mathbf{R}(\ell,m)$ for the Irregular1 (left) and 
Irregular2 (right) configurations at 750MHz. As expected, the general shapes of the $(\ell, m)$ space distribution are similar 
for the two cases, but with a wider area covered in the $(\ell, m)$ space for  the Irregular 2 configuration due to the 
larger array sizes. The broad outline of the shapes in this figure are also similar to those in Fig.\ref{fig-R-reg}, 
but here the distribution is more smooth and uniform due to the more widely spread-out $(\ell, m)$ coverage in the irregular configurations.
The features at $m=120$ and 350 in Irregular 2 configuration are much less prominent than in the Regular 2 case.

Figure \ref{fig-Errvar-irre} shows the corresponding error covariance matrix in the $(\ell, m)$ basis. Here the 
regions of larer error are spread out more widely, but the error value at the maximum is much reduced when compared with the 
regular configurations.  The Irregular 1 case has smaller errors than the Irregular 2 case, as the 
baselines are more concentrated in the former case which helps reducing the errors.

\begin{figure}
\centering
\includegraphics [width=0.7\textwidth] {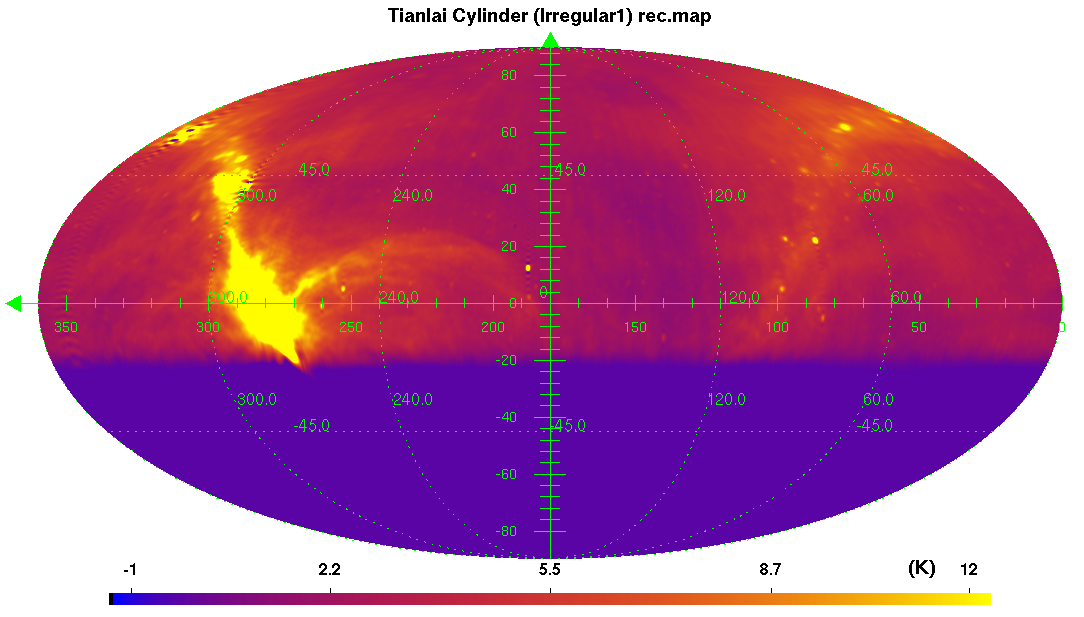}\\
\includegraphics [width=0.7\textwidth] {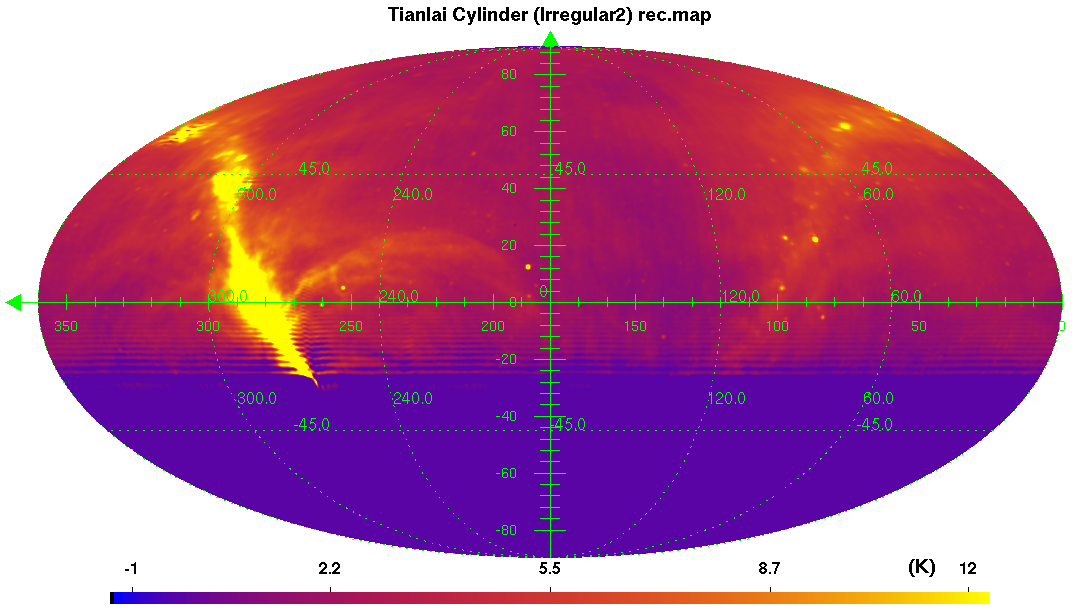}
\caption{Reconstructed sky map for the Irregular 1 (top) and Irregular 2 (bottom) configurations at 750 MHz. }
\label{fig-recmap-irregular}
\end{figure}

Figure \ref{fig-recmap-irregular} shows the simulated reconstruction map at 750 MHz with the  Irregular 1 (top) and 
Irregular 2 (bottom) configurations. We can see that in both cases, the reconstruction works relatively well, the spurious features
shown in Fig.\ref{fig-recmap-regular} are absent in these figures, and most features of the original map are well produced. There are still
some regions where the reconstruction shows some artefacts, e.g. the stripes at $(350^\circ, 60^\circ)$ and $(190^\circ, 12^\circ)$ in the 
Irregular  1 map, and the stripes south of the equator in the Irregular 2 map. However, the overall quality for the two maps are good.

\begin{figure}
\centering
\includegraphics[width=0.8\textwidth] {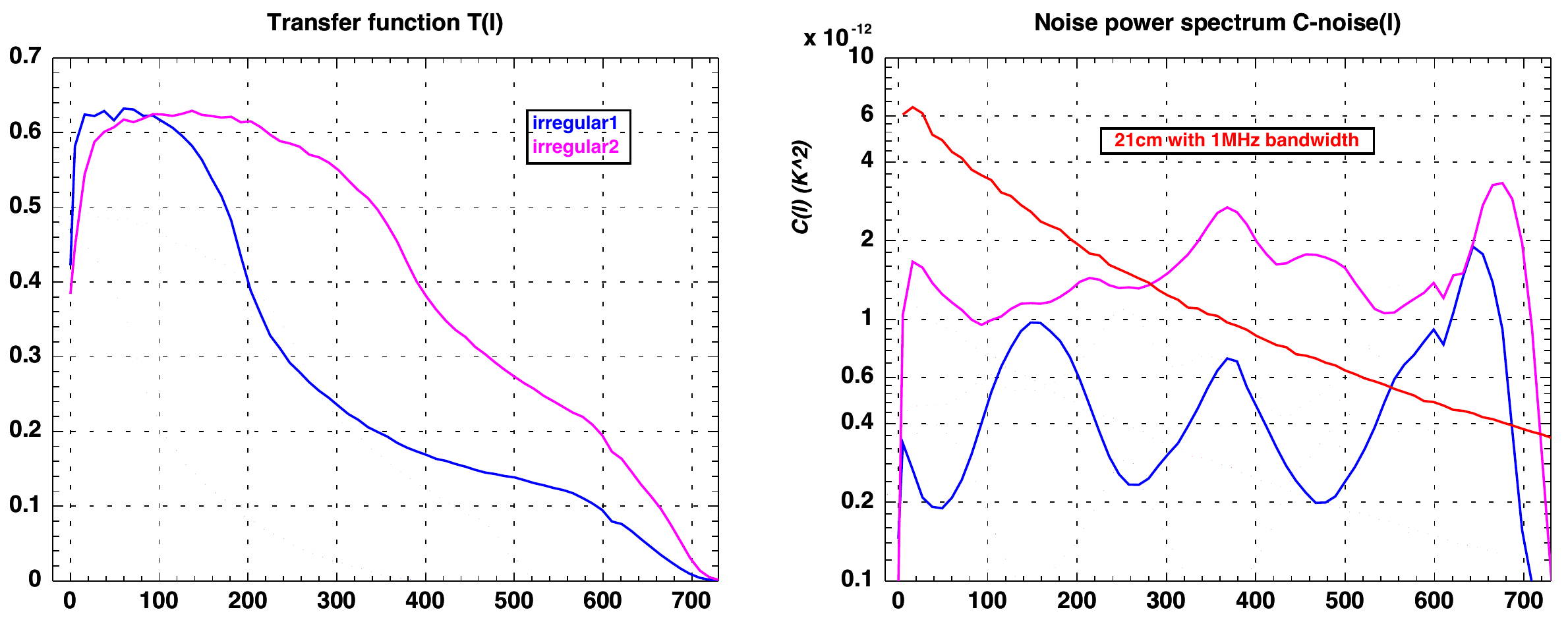}
\caption{Comparison of the transfer function $T(\ell)$ (left panel) and the noise power spectrum $C^\mathrm{noise}(\ell)$ (right panel) 
for the Irregular 1 and Irregular 2 configurations. }
\label{fig-tlcl-irre}
\end{figure}

In Figure \ref{fig-tlcl-irre}, we plot the power spectrum transfer functions $T(\ell)$ (left panel) 
and the noise power spectrum (right panel) for the Irregular 1  and Irregular 2 configurations. Here we have 
masked out the border pixels outside the band $ 0^\circ < \theta < 105^\circ $ which are not well constructed, 
and suppressed $(\ell, m)$ modes with large errors by applying a weight proportional $\sigma^{-2}_\mathcal{I}(\ell, m) $ to 
all modes which have error larger than
$ K  \sigma^2_{min} $, where $\sigma_{min}$ is the minumum value of the noise covariance matrix, and for the threshold value we choose
$K=50$. The transfer function decreases
toward higher $\ell$, but it is generally smooth, though there are curvatures at certain $\ell$s. The Irregular 1 configuration has higher 
response at lower $\ell$, but decreases more rapidly at higher $\ell$ as expected, as its baselines are concentrated in smaller regions and 
sensitive more to the larger angular scales. For the noise power spectrum, we see that the Irregular 1 configuration achieved lower noise power 
than the Irregular 2 configuration. In both cases the noise power spectrum show several peaks and troughs, 
which are due to the different density of baselines on the $(\ell, m)$ plane. We also draw the expected large scale structure 
21cm signal power on the same plot, where we assume cosmology from \citep{2014A&A...571A..16P}, and for 
neutral hydrogen we adopt $\Omega _{HI} b = 0.62 \times 10^{-3}$  \citep{2013MNRAS.434L..46S}. 
 The 21cm signal is only a few times
 the noise. Note that this is for the detection at a single frequency, we will have more frequency data, but at the same time there are 
 also complications of foreground removal and calibration, which are beyond the scope of the present work. Considering these factors, 
 we see that detecting the 21cm signal would be a great challenge.

\section{Conclusion}
\label{sec-conclusion}
The Tianlai experiment aims to make a low angular resolution, large sky area transit survey of the large scale structure 
by observing the redshifted 21cm line from neutral hydrogens. By adopting the transit survey strategy, where the telescope is fixed on the 
ground and scans the whole observable part of sky by the rotation of sky, the cost for building the telescope is reduced, and the 
instrument is also more stable. The transit scan is however very different from the tracking observation, and for the whole sky survey one
must also take into account the sphericity of the sky. 
 
We have developed an efficient, flexible and parallel code to make sky map from transit visibilities based on spherical harmonics 
transformations, which is applicable to any transit-type interferometer. This paper is the second of a series of papers presenting
our transit array data processing method. In this paper, we have applied this software to the simulation of map-making process for 
the Tianlai cylinder array pathfinder. In the simulation, we first compute the visibility time streams for several instrument 
configurations and scan strategies, and then reconstructed sky maps from these visibilities. The feed response and array 
geometry are assumed to be known and fully calibrated. 

The Tianlai pathfinder have 96 receiver feeds in total, averaging 32 on each cylinder. The cylinders 
could host about twice that amount of receiver feeds, 
leaving room for upgrading in the future after the present hardware designed is checked out throughly in experiment. 
We considered two types of arrangement of the feeds. In one type, the feeds are spaced at about one wavelength, which
covers less than half of the cylinder length with the 32 feeds on each. In the other type, 3/5 of the cylinder length is covered, 
with spacing of twice the wavelength.
 
On each cylinder the receiver feeds form regularly spaced linear arrays, which have grating lobes if the unit 
spacing is larger than half wavelength. Coupled with the large instantaneous field of view for the cylinder, 
the grating lobes could be a great obstacle for map-making. To solve this problem, irregular spacing can be introduced. 
A logistically simple solution is to adopt slightly different unit spacings on the three cylinders, but on each cylinder the spacing is still
uniform. We considered the arrangement of 31, 32, and 33 feeds on the three cylinders. 
With such irregular configurations, the grating lobes are reduced to very low level, and map reconstruction quality is enhanced.

We analysed the beams produced by the cylinders, and found that the features in the response matrix and noise variance matrix 
can be understood from these. We also computed the transfer function, and reconstructed map for both the Regular and Irregular 
instrument configurations.  We also computed the noise angular power spectrum, which determines the array sensitivity 
for cosmological 21 cm signal measurement. 
This may be regarded as a simplification of the real case, where the system temperature is dominated by the foreground radiation. 
We found that for a system temperature of 50 K, the 21cm angular power spectrum is a few times of the noise power in a single 1MHz 
narrow band.  Detecting such signal would be a difficult challenge, but the signal may be enhanced by considering joint measurement of 
the power spectrum over many spectral bins. In the present paper we study primarily the foreground map-making 
process, the calibration, foreground removal and 21cm signal extraction will be investigated in subsequent works.

\section*{Acknowledgements}
We would like to thank Ue-Li Pen and Richard Shaw for discussions. 
The Tianlai project is supported by the MoST 863 program grant 2012AA121701 and the CAS Repair and Procurement grant. 
PAON4 project is supported by PNCG,  Observatoire de Paris, Irfu/CEA and LAL/CNRS. 
JZ was supported by China Scholarship Council. XC is supported by the CAS strategic Priority Research Program XDB09020301, 
and NSFC grant 11373030. FW is supported by NSFC grant 11473044.

\bibliographystyle{raa}
\bibliography{jmapcylinder}

\end{document}